\documentclass[pra,twocolumn,
amsmath,amssymb,nofootinbib]{revtex4}

\usepackage{epsfig}
\usepackage{color}
\usepackage{graphicx}
\usepackage{bm}
\usepackage{mathtools}
\usepackage{gensymb}
\usepackage{ulem}
\usepackage{amsmath}
\usepackage{amssymb}

\usepackage{fontenc}  

 \usepackage[colorlinks,bookmarks=false,citecolor=blue,linkcolor=red,urlcolor=blue]{hyperref}

\begin{document}

 \title{Contextual unification of classical and quantum physics}

\author{Mathias Van Den Bossche$^1$ and Philippe Grangier$^2$}

\affiliation{$^1$ Thales Alenia Space, 26, avenue J.-F. Champollion, 31037 Toulouse, France}
\affiliation{$^2$ Laboratoire Charles Fabry, IOGS, CNRS, 
Universit\'e Paris Saclay, F91127 Palaiseau, France.}

\date{\today}

\begin{abstract}
Following an article by John von Neumann on infinite tensor products, we develop the idea that the usual formalism of quantum mechanics, associated with unitary equivalence of representations, stops working when countable infinities of particles (or degrees of freedom)  are encountered. This is because the dimension of the corresponding Hilbert space becomes  uncountably infinite, leading to the loss of unitary equivalence, and to sectorization. By interpreting physically this mathematical fact, we show that it provides a natural way to describe the ``Heisenberg cut'', as well as a unified mathematical model including both quantum and classical physics, appearing as required incommensurable facets in the description of nature. 
\end{abstract}

\maketitle

\section{Introduction}
\label{intro}

Despite its formidable predictability performance, quantum theory still appears to many as the realm of mysteries. This is the case especially when one tries to figure out ``what is going on" in a quantum phenomenon, which is highly exotic to our classical-shaped minds. The first solution to avoid getting lost in interpretation issues has been to ban any such questions. This has been the mainstream attitude since the first decades of quantum theory. The drawback of this solution is to strenghthen the mysterious aspect of quantum physics, and this turns out counterproductive, in terms of e.g. developing a physical intuition of quantum phenomena at stake in the emerging quantum technologies. The difficult situation quantum physicists face is actually not to avoid asking questions, but rather to find a way for asking the relevant ones. This has revealed a hard task, on which progress has been slow, but now cumulates a number of relevant findings, including both no-go theorems (e.g. Kochen-Specker \cite{KochenSpecker} or Bell \cite{BellInequalities}), and more positive ones (e.g. Gleason \cite{GleasonTheorem}). Beyond the  quantum properties of superposition/indeterminism, correlations/entanglement, several high level questions can be summarised as 
\begin{itemize}
\item While everything is built out of quantum elements, some systems obey the laws of quantum physics and some, classical systems, apparently do not;

\item  Even though quantum physics seems more fundamental, classical systems `embed' quantum systems as if the classical ones were necessary for the quantum ones to  be probed, in order to check the quantum predictions;

\item When it comes to retrieving and exploiting information from a quantum system, the device that is used 
is always a classical system;  

\item In particular, there is always a classical system (measurement device, environment...), larger  than 
the quantum system under consideration, and where the measurement results appear to be fundamentally random.
\end{itemize}
Depending on the mindset of the reader, these considerations may appear to have a strong ``Copenhagen" flavour, or more simply to be obvious from empirical evidence. Actually, they 
can be traced to the assumption of a cut in the description of physical systems, below which quantum laws apply and above which the laws of classical physics prevail. Following the literature we will call this the Heisenberg cut \cite{Hcut}.

The purpose of this paper is to propose a way to address this cut, by relying on a seminal, but surprisingly underrated, article by John von Neumann \cite{vonNeumann1939}, that explores the properties of the huge mathematical spaces built by infinite tensor products of usual Hilbert spaces\footnote{The concepts introduced by this paper by von Neumann have been generalised to give birth to the domain of operator algebras. Although the latter are richer, the approach of ref. \cite{vonNeumann1939} is sufficient for our present discussion of quantum macroscopic systems.}. These very large Hilbert spaces are typically the ones expected to describe the quantum properties of macroscopic systems -- possibly the ones that stand above the Heisenberg cut. In particular, we would like to show that the formalism of quantum physics, naturally and  suitably extended  to take into account the properties of these very large Hilbert spaces,  does include the Heisenberg cut.

We will also relate  these arguments to the approach known as ``Contexts, Systems and Modalities'' (CSM) \cite{CO2002,csm1,random,generic,csm4b,csm4c,grangierCompleting,ContextualInference,myst,context}, that proposes an ontological clarification of quantum physics in a contextually objective framework  \cite{CO2002}, reminiscent of the Copenhagen views.  As we will see, there is an excellent fit between CSM and the arguments put forward in the present work.

This article  is organised the following way. Section \ref{sec_infiniteTensorProducts} translates the main elements of von Neumann's paper on infinite tensor products into the mathematical language currently used in quantum mechanics (QM), to help understanding by contemporary physicists\footnote{We mean low-energy physicists, a full quantum field theory presentation is outside the scope of the present article. We note also that a non-mathematically-minded reader may skip Section \ref{sec_infiniteTensorProducts} in a first read, and to come back to it later on, bearing in mind that the most important result of this section is the Hilbert space breakdown theorem.}. Section \ref{sec_largeSystems} derives general consequences of these results  in terms of large quantum systems properties and measurement processes, and Section \ref{sec_examples} illustrates and discusses these consequences on two examples, using the usual  QM  language. 
Finally Section \ref{sec_CSM_wrap_up} shows how to exploit these mathematical ideas in the physical CSM framework.

\section{Infinite tensor products}

\label{sec_infiniteTensorProducts}

{\color{black} This} paper addresses explicitly the quantum description of large systems. We will call a large system a system made of a large number $N$ of elementary quantum systems $S_\alpha$ (degrees of freedom), possibly interacting with one another, each described by a separable Hilbert space $\mathcal{H}_\alpha$. As is usually done in physics, we will look for emerging properties at the limit $N\rightarrow \infty$. The space of states of the large system is expected to be the generalisation of the tensor product of all Hilbert spaces of the elementary systems it is made of.  The results of \cite{vonNeumann1939} are valid both for countably and uncountably infinite $N$, even though the examples in the subsequent sections consider essentially the countably infinite case. In both cases one should be aware that some peculiarities are likely to emerge, as the following example shows. 

Taking the simplest case of $N$ elementary 2-state systems, one can expect that the limit of the $N \rightarrow \infty$  tensor products of its ($N \sim \aleph_0$) Hilbert spaces, is a space of states of non-countable dimension $\aleph_1 = 2^{\aleph_0}$, where $\aleph_0$ and $\aleph_1$ are Cantor's usual notations. A vector space with non-countable dimension is in general non-separable, and this can be the reason for new, richer than usual properties to show up, as compared to the state spaces considered in usual quantum physics, that are separable \cite{StreaterWightman}. It is this kind of properties that this section will introduce. 

As we want to make statements on the issues of measurements on these large quantum systems, we need to have a way to compute probability amplitudes. This means that we need to know how to project a state onto another one. Following von Neumann, we want to generalise the idea that the scalar product of two states that live in the tensor product of $N \rightarrow \infty$ Hilbert spaces is the complex product of the $N\rightarrow \infty$ scalar products in each Hilbert space. We therefore need first to define the convergence of infinite products of complex numbers before going to the construction of infinite tensor products. We finally then describe some specific properties of the space generated by infinite tensor products.
The sections below present a short synthesis of \cite{vonNeumann1939}, and the proofs of the results given here are to be found in this reference.

\subsection{Infinite complex products}

{\bf Convergence of infinite complex products} -- Let $I$ be an infinite set of indices (countable or uncountable) 
and let $\{z_\alpha\}_{\alpha\in I}$ be a set of complex numbers.  The convergence of the infinite product of the $z_\alpha$ is defined as:
\begin{eqnarray*} 
&\prod_{\alpha\in I} z_\alpha \mbox{ converges to value } Z \in \mathbb{C} \iff  \\
&\forall \varepsilon > 0,  
\exists \mbox{ a finite } J_\varepsilon\subset {\color{black} I}  \mbox{ such that } \\
&\forall J_\text{all} =\{ \alpha_1, ...,\alpha_N\}\subset {\color{black} I}  \mbox{ all } \alpha_i \mbox{'s different and }   
J_\varepsilon \subset J_\text{all}, \\
&\vert z_{\alpha_1} z_{\alpha_2} ... z_{\alpha_N} - Z \vert< \varepsilon. \qquad 
\end{eqnarray*}  
\vskip 2mm

\noindent {\bf Properties} --  One can prove the following properties for such infinite products.

\begin{itemize}

\item The value of $Z$ is unique if it exists. 

\item  If all $z_\alpha \in \mathbb{R}^{+}$, their product converges only if either $\sum_\alpha \max(z_\alpha-1, 0)$ converges or some $z_\alpha = 0$.  

\item  If $z_\alpha$ are real or non-real complex numbers, their product converges if and only if either 
$\prod_\alpha \vert  z_\alpha \vert  = 0 $ in the above sense{\color{black}, or } 
$0<\prod_\alpha \vert  z_\alpha \vert   <\infty$ and $\sum_\alpha \arg(z_\alpha) <\infty$.

\item In case of discrete, ordered infinite products, this definition is equivalent to the convergence of the sum of the logarithm{\color{black}s of the $z_\alpha$,} at least above a given rank where the logarithm remains on the same branch.
\end{itemize}

To handle infinite products that converge in modulus but not in argument, {\it quasi-convergence} is defined as follows. If 
$\prod_\alpha \vert z_\alpha\vert $ converges then $\prod_\alpha z_\alpha$ is quasi-convergent and, either $\prod_\alpha z_\alpha$ converges to its value, or its argument is too unstable for $\prod_\alpha z_\alpha$ to converge, and the quasi-convergence value is set arbitrarily to 0. ``Convergence'' is to be understood as ``quasi-convergence'' below, unless stated otherwise.

\subsection{Infinite tensor products} 

It takes introducing a few more concepts before the infinite tensor product of Hilbert spaces can be defined. One wants to introduce infinite tensor products of Hilbert spaces so as to be able to compute quantum probability amplitudes. This means that a scalar product, and thus legitimate linear forms have to be introduced. 
Our presentation thus starts with these objects, which belong to the dual of the space we target, and then takes the dual and completes it topologically, to construct a full-fledged Hilbert space that is 
unique, metrised, and complete.  
\\

\noindent {\bf Convergent sequences} (Von Neumann's $C$-sequences) -- As above, let $I$ be an infinite set of indices (countable or uncountable) and let 
$\{\mathcal{H}_\alpha\}_{\alpha\in I}$ 
be a set of (separable) Hilbert spaces attached to each $\alpha$. 

Let 
$ \{ \vert  \phi_\alpha \rangle \}_{\alpha \in I}$
with each 
$\vert \phi_\alpha\rangle \in\mathcal{H}_\alpha$\footnote{Note that this requires the axiom of choice.}. 
Then $\{\vert \phi_\alpha\rangle\}_{\alpha\in I}$  is a {\it convergent-sequence} $\iff   \prod_\alpha  \vert \vert \mbox{ } \vert \phi_\alpha\rangle\mbox{ } \vert \vert $ converges.

\noindent {\bf Non-trivial convergent-sequences}(Von Neumann's $C_0$-sequences) -- They  are defined as the convergent-sequences $\{ \vert \phi_\alpha\rangle\}_{\alpha\in I}$ for which $\sum_\alpha \vert  \sqrt{\langle \phi_\alpha \vert  \phi_\alpha\rangle} - 1 \vert  $ converges (equivalently, without the root). It will appear below that they are the controlled and non-vanishing infinite tensor products. These well-behaved, non-trivial elements of the infinite tensor product space will be the ones of physical interest.
\vskip 2mm

\noindent {\bf Elementary linear forms} --  Based on the property that {\color{black} $\prod_\alpha \mbox{ } \langle\psi_\alpha\vert \phi_\alpha\rangle\mbox{ }$ quasi-converges if $\{\vert \phi_\alpha\rangle\}_{\alpha\in I}$ and $\{\vert \psi_\alpha\rangle\}_{\alpha\in I}$ are two convergent-sequences,} one can define elementary linear forms on the convergent-sequences of the cartesian product of the $\mathcal{H}_\alpha$'s. For instance 
\begin{equation}\Phi : \{\vert \psi_\alpha\rangle\}_{\alpha\in I} \rightarrow \prod_{\alpha \in I} \mbox{ } \langle\phi_\alpha\vert \psi_\alpha\rangle\mbox{ }\end{equation} 
\noindent {\bf Composite linear forms} --  By considering the linear combination of these elementary linear forms, one can define some linear forms on a set that is the linear combination of the above cartesian product. For instance, with $m \in \{1,...,M\}$, the $M$ convergent-sequences $ \{\vert \phi_\alpha^m\rangle\}_{\alpha\in I}$ and $M$ complex coefficients ${\it \Phi}_m$ can be used to build more linear forms such as  

\begin{equation}
\Phi : \{\vert \psi_\alpha\rangle\}_{\alpha\in I} \rightarrow \sum_{m=1}^M {\it \Phi}_m^* \prod_{\alpha \in I} \mbox{ } \langle\phi_\alpha^m\vert \psi_\alpha\rangle\mbox{ .}
\end{equation}
We will {\it note} it $\langle \Phi\vert  = \sum_{m=1}^M {\it \Phi}_m^* \otimes_{\alpha \in I}\langle \phi_\alpha^m\vert $. 
\vskip 2mm

\noindent {\bf Scalar product of linear forms} -- Let  $\{\vert \phi_\alpha^m\rangle\}_{\alpha\in I, m=1,M}$ and $\{\vert \psi_\alpha^n\rangle\}_{\alpha\in I, n=1,N}$ be two sets of convergent-sequences and define the associated linear forms $\langle \Phi\vert $ and $\langle \Psi\vert $ as 

\begin{equation}
\langle\Phi\vert  = \sum_{m=1}^M {\it \Phi}_m^* \otimes_{\alpha \in I}\langle \phi_\alpha^m\vert ,\langle\Psi\vert  = \sum_{n=1}^N {\it \Psi}_n^* \otimes_{\alpha \in I}\langle\psi_\alpha^n\vert ,
\end{equation}
then one can define their scalar product as 

\begin{equation}
\langle\Phi\vert \Psi\rangle 
:= \sum_{m=1}^M \sum_{n=1}^N 
{\it \Phi}_m^* {\it \Psi}_n \prod_{\alpha \in I} \langle \phi_\alpha^m \vert  \psi_\alpha^n\rangle
\end{equation}
that can be shown to have all the 
properties of a hermitian scalar product. 
\vskip 2mm

\noindent {\bf Complete infinite tensor product space} -- $\mathcal{H}=\otimes_\alpha \mathcal{H}_\alpha$ is defined as the dual of the set of forms $\langle \Phi\vert $ such that there exists a sequence $\langle \Phi_1\vert ,...,\langle \Phi_i\vert ,...$ such that 
\\
a) for all convergent sequences $\{ \vert \psi_\alpha\rangle\}_{\alpha\in I}, \Phi (\{\vert \psi_\alpha\rangle\}) = \lim_{i\rightarrow \infty} \Phi_i (\{\vert \psi_\alpha\rangle\})$   
\\
b) $\lim_{i,j\rightarrow \infty} \vert \Phi_j(\{\vert \psi_\alpha\rangle\})-\Phi_i(\{\vert \psi_\alpha\rangle\})\vert  =0 $     

\noindent This set has the following properties.
\begin{itemize}
\item $\mathcal{H}$ includes the dual of the space of the above composite linear forms and generalises the notion of tensor product to an infinite number of  Hilbert spaces.
 
\item The above composite linear forms on that space provide tools to deal with the elements of $\mathcal{H}$. 

\item There could be more to the infinite product $\mathcal{H}$ than the duals of the above linear forms, built on convergent sequences. We can at least make statements about the part made from {\color{black} these,} 
which provide workable objects. 

\item An element of $\mathcal{H}$ built with the elements of the sequence $\{\vert \phi_\alpha\rangle\}_{\alpha\in I}$ with each $\vert \phi_\alpha\rangle \in \mathcal{H}_\alpha$ will be noted $\vert \Phi\rangle=\otimes_{\alpha \in I} \vert \phi_\alpha\rangle$.  
\end{itemize}

\noindent {\bf Hilbert space Theorem}(von Neumann's theorem II) --  The above space, generated by taking the dual of the limit completion of a finite sum of infinite direct products of vectors, equiped with the above scalar product, is a Hilbert space.  

\noindent In other words, although in the first place, only the convergent-sequence part of $\mathcal{H}$ is well behaved enough to be equipped with the above scalar product,  as long as an infinite tensor product includes up to the finite sum of infinite products built with convergent sequences, the scalar product will exist, be unique, and again have all the usual properties of an Hermitian scalar product. 

\noindent {\bf Scalar product} -- In order to define it on the complete infinite tensor product space, let  

\begin{equation}
\vert \Phi\rangle = \lim_{j\rightarrow\infty} \vert \Phi_j\rangle \mbox{ with }\vert \Phi_j\rangle = \sum_{m=1}^M {\it \Phi}_m^j \otimes_{\alpha\in I} \vert \phi_\alpha^m\rangle
\end{equation} 
and
\begin{equation}\vert 
\Psi\rangle = \lim_{j\rightarrow\infty} \vert \Psi_j\rangle\mbox{ with }\vert \Psi_j\rangle = \sum_{m=1}^N {\it \Psi}_m^j \otimes_{\alpha\in I} \vert \psi_\alpha^m\rangle
\end{equation} 
where 
\begin{equation}
\{\vert \phi_\alpha^m\rangle\}_{\alpha\in I, m=1,M}\mbox{ and }\{\vert \psi_\alpha^m\rangle\}_{\alpha\in I, m=1,N}
\end{equation} 
are two sets of convergent sequences. Then  
\begin{equation}
\langle \Phi\vert \Psi\rangle :=\lim_{j\rightarrow\infty} \langle \Phi_j\vert \Psi_j\rangle
\end{equation} 
exists, is unique, has the properties of a Hermitian scalar products and verifies Schwarz's inequality.

\noindent {\bf  Topology of} $\mathcal{H}$ -- $\mathcal{H}$ can be metrised by a distance defined as $d(\vert \Phi\rangle,\vert \Psi\rangle)= (\langle\Phi\vert -\langle\Psi\vert )(\vert \Phi\rangle-\vert \Psi\rangle)$. With this distance, it can be shown that the space of finite sums of infinite tensor products built out of convergent sequences is dense in $\mathcal{H}$ which is topologically complete.  

\noindent {\bf Uniqueness theorem} (von Neumann's Theorem III) -- $\mathcal{H}$ is a Hilbert space equiped with a distance and is unique up to an isomorphism.

\subsection{Hilbert space breakdown into equivalence sectors} 

\noindent {\bf  Equivalence sectors of non trivial convergent-sequences.} One can show that the relation between two non-trivial convergent-sequences defined by  
\begin{equation} 
\{ \vert \phi_\alpha\rangle\}_{\alpha\in I} \sim \{ \vert \psi_\alpha\rangle\}_{\alpha\in I} \iff \sum_{\alpha \in I} \vert \langle \phi_\alpha\vert \psi_\alpha \rangle -1\vert  \mbox{ converges}
\end{equation} 
is an equivalence relation. We will call {\it sectors} its equivalence classes. 
\vskip 2 mm

\noindent {\bf Normed representative theorem} of an equivalence sector. It can be shown that each equivalence sector contains at least one non-trivial convergent sequence $\{ \vert \phi_\alpha\rangle\}_{\alpha\in I}$ such that $\forall \alpha \in I, \langle \phi_\alpha \vert  \phi_\alpha\rangle = 1 $  
\vskip 2 mm

\noindent {\bf Breakdown theorem} (Von Neumann's theorem I)
\\ 
a) If two non-trivial convergent-sequences are in two different sectors, then they are orthogonal, i.e. one has  $(\otimes_\alpha \langle \phi_\alpha\vert ) (\otimes_\alpha\vert \psi_\alpha\rangle) = 0$.

\noindent b) If $(\otimes_\alpha \langle \phi_\alpha\vert ) (\otimes_\alpha\vert \psi_\alpha\rangle) = 0$ and the two non-trivial convergent-sequences are in the same sector, then $\exists \alpha \in I, \langle \phi_\alpha\vert  \psi_\alpha \rangle = 0$

\noindent  c) If $\# \{\alpha \in I, \vert \psi_\alpha\rangle \neq \vert \phi_\alpha\rangle\} < \infty$ 
then $\{ \vert \phi_\alpha\rangle\}_{\alpha\in I} \sim \{ \vert \psi_\alpha\rangle\}_{\alpha\in I}$
\vskip 2 mm

\noindent {\bf   Incomplete tensor product} -- If  $\mathcal{C}$ is a sector of 
$\otimes_{\alpha \in I} \mathcal{H}_\alpha$, the set defined by the non-trivial convergent sequences of $\mathcal{C}$
\begin{equation}
\otimes_{\alpha \in I}^\mathcal{C} \mathcal{H}_\alpha := \{ \otimes_{\alpha \in I} \vert \phi_\alpha\rangle, \{\vert \phi_\alpha\rangle\}_{\alpha \in I} \in \mathcal{C}\}
\end{equation}
is called the {\it incomplete infinite tensor product} of the $\mathcal{H}_\alpha$'s over $\mathcal{C}$.
\vskip 2 mm

\noindent {\bf Direct sum theorem} -- The incomplete tensor products are in direct sum, and their whole set (taking all equivalence sectors) generates the complete infinite tensor product space. 
\vskip 2 mm

\noindent {\bf Finite change theorem} -- Let $\{ \vert \phi_\alpha^0\rangle\}_{\alpha \in I}$ be a non-trivial convergent sequence of $\mathcal{C}$ with $\forall \alpha\in I, \langle \phi_\alpha^0 \vert  \phi_\alpha^0\rangle = 1$. Then $\mathcal{C}$ is the set of all non-trivial convergent sequences $\{ \vert \phi_\alpha\rangle\}_{\alpha \in I}$ where $\vert \phi_\alpha\rangle \neq \vert \phi_\alpha^0\rangle$ for a finite number of $\alpha$'s only.   

The following two lemmata can be derived from von Neumann's article and are introduced here as they will bring some clarity to the understanding of the Heisenberg cut. Their proofs are given in the Annex A1.
 \vskip 2 mm
 
\noindent {\bf Vector sectorisation limit lemma} --  
Let $\vert \Phi\rangle = \otimes_{\alpha \in I} \vert \phi_\alpha\rangle$ and $\vert \Psi\rangle = \otimes_{\alpha \in I} \vert \psi_\alpha\rangle$ where $\{ \vert \phi_\alpha \rangle \} _{\alpha\in I}$ and $\{ \vert \psi_\alpha \rangle \} _{\alpha\in I}$ 
are two non trivial convergent sequences of two distinct sectors of $\mathcal{H} = \otimes_{\alpha\in I}\mathcal{H}_\alpha$. 
Let $I_N =\{ \alpha_1, \alpha_2,...,\alpha_N \} \subset I$, all distinct, with $N\in \mathbb{N}$, $\vert \Phi_N\rangle = \otimes_{\alpha \in I_N} \vert \phi_\alpha\rangle$ and $\vert \Psi_N\rangle = \otimes_{\alpha \in I_N} \vert \psi_\alpha\rangle$, then 
\begin{equation}\forall \varepsilon > 0, \exists N \in \mathbb{N}, \vert \langle \Psi_N\vert \Phi_N\rangle\vert  < \varepsilon
\end{equation}
\vskip 2 mm

\noindent {\bf Operator sectorisation limit lemma} --  
Let $\vert \Psi\rangle = \otimes_{\alpha \in I} \vert \psi_\alpha\rangle$ 
{\color{black} and $\vert \Phi\rangle = \otimes_{\alpha \in I} \vert \phi_\alpha\rangle$}  
where $\{ \vert \psi_\alpha \rangle \} _{\alpha\in I}$ 
{\color{black} and $\{ \vert \phi_\alpha \rangle \} _{\alpha\in I}$ are two} 
non-trivial {\color{black}normed} convergent sequence{\color{black}s} in sector $\mathcal{C}$ of $\mathcal{H} = \otimes_{\alpha\in I}\mathcal{H}_\alpha$. 
Let $I_N =\{ \alpha_1, \alpha_2,...,\alpha_N \} \subset I$, all distinct, with $N\in \mathbb{N}$ and $\vert \Psi_N\rangle = \otimes_{\alpha \in I_N} \vert \psi_\alpha\rangle$ {\color{black}, $\vert \Phi_N\rangle = \otimes_{\alpha \in I_N} \vert \phi_\alpha\rangle$ 
}. Let {\color{black}$\hat A$ } be a bounded operator on $\mathcal{H}$ 
and let {\color{black}$\hat A_N$ }be the restriction of $\hat A$ to $\otimes_{\alpha \in I_N}\mathcal{H}_\alpha$.  
{\color{black} 
Let $p(\hat A) := \# \{\alpha, \vert \phi_\alpha\rangle \neq e^{i\xi_\alpha} \vert\psi_\alpha\rangle, \xi_\alpha \in \mathbb{R} \}$. 

\noindent a) If $p(\hat A) \l \infty$ then  $\langle \Phi \vert \hat A \vert \Psi\rangle$ can take any bounded value in $\mathbb{C}$ 

\noindent b) If $p(\hat A)$ infinite, then 
\begin{equation}
\forall \varepsilon > 0, \exists N \in \mathbb{N}, \langle \Phi_N\vert \hat A_N\vert \Psi_N\rangle < \varepsilon
\end{equation}
}
\noindent In other words, when $N\rightarrow \infty$ the change of $\vert \Psi\rangle$ under {\color{black}$\hat{A}$} bounded (all the more, unitary) is orthogonal to $\vert \Psi\rangle$ as soon as {\color{black}$\hat{A}$}  modifies more than a finite number of terms in the infinite tensor product.  Intuitively, even an infinitesimal rotation in the full $\mathcal{H} = \otimes_{\alpha\in I}\mathcal{H}_\alpha$ leads to an orthogonal state in the general case. Only remaining in the same sector {\color{black}-- i.e. acting on a finite number of $\alpha$'s -- } allows keeping usual properties of rotations.

\section{Quantum states of large systems} 
\label{sec_largeSystems}

In this section, we draw the first physical consequences of the mathematical considerations of the previous section. We address first the properties of the states of large quantum systems that can be deduced from the above results. We then consider the special case of measurements. Finally, and very importantly, we discuss the room that is left for quantum phenomena to occur. 

\subsection{Macroscopic system quantum state properties} 

The formalism summarised in section II applied to 
usual quantum physics allows understanding under what conditions a large system can be in a superposed state and thus have quantum properties. In this section, we discuss the properties of the quantum states of macroscopic systems, in particular superposition of macroscopically different quantum states. In concrete experiments, there are two ways to obtain a system in superposed state: by evolution or by preparation through measurement. We examine both situations in turn below.
\vskip 2 mm

\noindent {\bf Evolution.} A system evolves to a superposed state when it is initially prepared in a state that is not an eigenstate of the  Hamiltonian of the closed system in which it participates. A simple example are Rabi oscillations between two states that are not eigenstates of the system Hamiltonian. During Rabi oscillations, the system is, in general, in a superposed state of the two elementary states under consideration. A more complex example is the de-excitation of an atom by random emission of a photon. At any time, before the photon is detected, the atom and the quantised electromagnetic field are in the superposition of a state with an excited atom and no photon, and a state with the atom in its ground state and an emitted photon. In this case too, neither states are eigenstates of the atom and quantised electromagnetic-field closed-system Hamiltonian.
\vskip 2 mm

\noindent {\bf Measurement.} A system can be prepared in a superposed state  of two values of an observable described by operator $\hat A$ by performing a projective measurement of a different observable which associated operator $\hat B$ does not commute with $\hat A$. A simple example of this is to prepare the polarisation state of a photon in the superposition of a horizontal state and a vertical state, by measuring its polarisation along a direction that is at a non-vanishing angle with the vertical and horizontal directions. The question whether measurements are a particular case of evolution is left for later.  
\vskip 5 mm

\noindent {\bf Large systems.} Let us now consider large systems with $N$ microscopic subsystems. With the usual assumptions of quantum mechanics, the Hilbert space of the macroscopic system is the tensor product $\mathcal{H} = \otimes_{i=1}^N \mathcal{H}_i$, where the microscopic subsystems Hilbert spaces $\mathcal{H}_i$ have a countable dimension $d_i = \dim \mathcal{H}_i$. This choice is justified by the fact that quantum systems are systems where there is always an elementary building block that has at most a countable (and frequently, finite) number of states. As is usual in statistical physics, we will describe a macroscopic system as the system which behaves as the limit when $N\rightarrow \infty$ of the behaviour of the $N$-subsystem large system. 
The properties of the above sectors transpose in physics into a number of statements.
\begin{itemize}
\item There is always a macroscopic state that can be built with the product of normed microscopic states, and so the macroscopic state is well defined [Normed representative theorem].    
\vskip 3 mm
\item Macroscopic state spaces separate into sectors {\color{black}whose} states differ by  a macroscopic fraction (i.e. which ratio to $N$ remains finite when $N\rightarrow \infty$) of the microscopic states [Breakdown theorem, c)]. These sectors are the {\it classical sectors} of the macroscopic system state space.    
\vskip 3 mm
\item Macroscopically different macroscopic states are orthogonal [Breakdown theorem, a)]
\vskip 3 mm
\item The evolution of a macroscopic system under a Hamiltonian which affects only a finite number (i.e. a microscopic fraction) of elementary quantum subsystems leaves the systems in the same macroscopic  state as it does not change sectors (Operator sectorisation limit lemma, a, applied to evolution operators {\color{black} that are unitary, thus bounded}).
\vskip 3 mm
\item The evolution of a macroscopic system under a Hamiltonian that affects a macroscopic fraction of elementary quantum subsystems, and of which the macroscopic state is not an eigenstate, takes the macroscopic state of the system through states that are all orthogonal {\color{black} with one another} at each instant, as soon as the evolution operator is not the identity (Operator sectorisation limit lemma, {\color{black}b}, applied to evolution operators).

\end{itemize}

\noindent {\bf Superposition of large system states.} 

\begin{itemize}
\item Preparing a superposed state of macroscopic system $S$ initially in state $\vert \Psi\rangle$ could be carried out by applying a measurement procedure that projects  $\vert \Psi\rangle$, on a target state $\vert \Phi\rangle$ that differs by a macroscopic fraction of the microscopic element states. But with these assumptions, $\vert \Psi\rangle$ and $\vert \Phi\rangle$ belong to two different sectors, so the macroscopic states will be orthogonal, and thus the probability amplitude to succeed vanishes.
\\
\item Letting a macroscopic system evolve towards a superposed state of a macroscopic fraction of its elements under the action of an inter-sector non-diagonal Hamiltonian $\hat H$ will not succeed either, because starting from state $\vert \Psi\rangle$ at $t=0$, all the successive states 
\begin{equation}
\hat U(t)\vert \Psi\rangle = e^{i\hat H t}\vert \Psi\rangle
\end{equation}
will be orthogonal to $\vert \Psi\rangle$ {\color{black} (or to any $\vert \Phi\rangle$ in the same macroscopic sector as $\vert \Psi\rangle$)} as soon as $\hat U(t) \neq \hat I$, i.e. $t\neq 0$, by the Operator sectorisation limit lemma. This leaves no room for a non-vanishing overlap needed to build a macroscopic superposition of different macroscopic states.
\end{itemize}

{\color{black} So unless another experimental procedure that would escape these constraints can be proposed,
one has to accept that the internal logic of the usual quantum physics framework, pushed to the limit of large systems as described by \cite{vonNeumann1939}, naturally implies that {\it superposition of macroscopically different quantum states cannot be produced, and thus do not exist}. }

{\color{black}In addition, }superposition being the way to describe quantum measurement indeterminism, this {\color{black}also} means that large-system measurements are deterministic, i.e. they abide by the classical physics laws. The example given in section \ref{sec_examples} below illustrates the mechanism of the separation to clarify the way it occurs.

{\color{black} 
\subsection{Density operators and self-decoherence of large systems.} 
As a consequence of the above results, the density operator of a macroscopic system in a mixed state in set $S = \{ \vert \Psi_i\rangle, i =1,...,s\}$ with probabilities $ \{p_i\}$ is built out of infinite tensor products and {\it not out of superpositions} of infinite tensor products that cannot be created, 
\begin{equation}
\hat \rho = \sum_{i=1}^s p_i \vert \Psi_i\rangle\langle \Psi_i\vert=\sum_{i=1}^s p_i \otimes_{\alpha \in I}\vert \psi^i_\alpha\rangle\langle \psi^i_\alpha\vert
\end{equation} 

This means that the density operators of large systems are block-diagonal, each block corresponding to a macroscopic sector. Before reaching the $ N\rightarrow \infty$ limit, in the large-but finite size case, the Operator sectorisation lemma shows that the off diagonal-block elements can be bounded by an arbitrarily small value taking large enough a system. In the language of operator algebras, $\hat \rho$ is built as the sum of infinite projectors (the $\otimes_{\alpha \in I}\vert \psi^i_\alpha\rangle\langle \psi^i_\alpha\vert $'s), and is thus a `state' of a type-III von Neumann algebra \cite{grangierCompleting}.

The fact that density operators of large systems are block diagonal within macroscopically equivalent sectors, with bounded off-diagonal elements in the large-but-finite size case, {\color{black}  provides an explicit  decoherence mechanism} 
where there is no need to lose information (i.e. trace on the) outside of the system to obtain a superselected density matrix, as the usual Zeh-Zurek decoherence theory does \cite{zehDecoherence,zurek,cr}. The structure of the Hilbert space provides this, exactly at the limit, and with as-low-as-desired residual probability amplitudes ($\l \varepsilon $) if the system is large enough. We propose to term this macroscopic, built-in effect {\it von Neumann decoherence.} 
{\color{black} A consequence of this algebraic structure is that superpositions of macroscopic states, the so-called Schr\"odinger's cat's states, are simply not supported by the formalism. This has clearly important consequences  for quantum measurements, that we will consider now.}
}
\subsection{{\color{black} Consequences on the measurement description}}
This section considers the {\color{black} situation} where the large quantum system is the combination of a microscopic {\color{black} quantum }system and a macroscopic measurement device, both described {\color{black} using the {\color{black}above} formalism}, and 
in mutual interaction. We will show now that the above considerations provide {\color{black} both a constraint on how to describe the measurement process, and a mechanism that leads to a superselected density operator for the microscopic system and measurement device,}
 {\color{black} taken together}.
\\

{\color{black} 
Let $S$ be a quantum system with Hilbert space $\mathcal{H}_S$ spanned by the basis formed by the eigenstates of a given observable $\{ \vert i\rangle \}_{i=1,M} $, and initially in the state $\vert s \rangle = \sum_{i=1,M} s_i \; \vert i \rangle $. The standard way to describe the measurement process of this observable is to decompose it in two steps: }
\vskip 2 mm

{\color{black} 
\noindent $\bullet$ In a first step the state of $S$  is entangled with a second system $A$ of ancillary degrees of freedom{\color{black}, designed } to separate the different components of the microscopic system state on different output channels that correspond to different values of the observable under consideration. Let $\{ \vert a_i\rangle \}_{i=1,M} $ be the states of these ancillary variables. They can be for instance the particle momentum states in a Stern-Gerlach experiment, or the wavevector state in a polarisation measurement. This step can be described as 
\begin{equation}
\vert s \rangle \otimes \vert a_0 \rangle = \sum_{i=1,M} s_i \vert i \rangle \otimes \vert a_0 \rangle \rightarrow \sum_{i=1,M} s_i \; \vert i \rangle \otimes \vert a_i\rangle
\end{equation}
where $ \vert a_0\rangle $ is the 
initial state of the anciliary degrees of freedom.  {\color{black} Denoting $\vert i \rangle \otimes \vert a_i \rangle$ as $\vert i, a_i \rangle$} the density operator of $S \cup A$ writes 
\begin{equation}
\hat \rho_{S\cup A} = \sum_{i,j=1,M} s_i s_j^*\vert i, a_i \rangle \langle j, a_j \vert
\end{equation}
{\color{black}  It is important to note that this evolution leads to a new superposition of states, and is allowed to be unitary because the number of ancillary degrees of freedom is finite. This first step is usually called the ``pre-measurement", and it corresponds to a reversible {\it analysis} of the state.}
\vskip 2 mm

\noindent $\bullet$ Then in a second step, {\color{black} the actual {\it detection process} is triggered} by involving one or several detectors $D_i$. For this, each different state $\vert a_i\rangle$ of the ancillary degrees of freedom starts interacting with a larger and larger number of states of other degrees of freedom in $D_i$. This gives rise to {\color{black} a ``macroscopic cascade'' or amplification process} that leads to macroscopic states of $S\cup A \cup D_i$ of the form $\vert i, a_i \rangle \otimes_{\alpha \in I} \vert \psi_\alpha^i\rangle$. 
%
All the skill of detector {\color{black}designers} resides in the way to create this macroscopic {\color{black} amplification process, as it will be illustrated on a concrete exemple in Sec. \ref{QND}}. At this stage, the density operator of $S\cup A \cup D_i$'s keeps growing, but as the number of degrees of freedom is very large {\color{black} and unbounded in practice}, it is legitimate as in §3.2 to consider it as well modeled by the $N \rightarrow \infty$ limit. 
 {\color{black} Within each $D_i$  the different states are macroscopically the same, and can be represented by one of them, $\vert {\cal D}_i\rangle$. We note that, since $\vert {\cal D}_i\rangle$ stands for a {\color{black} macroscopic sector} subspace of $D_i$'s infinite tensor product Hilbert space, it represents a block and not a single diagonal element.}

So the global density operator becomes block-diagonal,
one block per macroscopic sector (i.e. measurement {\color{black}outcome}) and we obtain a superselected density matrix
\begin{equation}
\hat \rho_{S\cup A\cup D}  = \sum _{i=1,M} \vert s_i \vert^2\vert i, a_i, {\cal D}_i \rangle \langle i, a_i, {\cal D}_i \vert
\end{equation}
}
{\color{black} 
Since this density matrix is diagonal, the $\vert s_i\vert ^2$'s can be interpreted as the classical probability for $S\cup A \cup D$ to be in state $\vert i, a_i, {\cal D}_i)$ and thus $S$ to be observed in state $\vert i\rangle$. After this observation the density matrix can be updated to the projector $\vert i, a_i, {\cal D}_i\rangle \langle i, a_i, {\cal D}_i\vert $ corresponding to the observed result.  
{\color{black} This last step is the  ``measurement actualisation'', and this}
situation reminds the one in statistical physics, where the state is not known for lack of information. However it is essential to keep in mind that here the randomness has a quantum origin, in the superposition of states, i.e. the non-determination of the observable value before the measurement; 
see also below. On the other hand, there is no need for a non-unitary projection postulate, since the result comes from the (non-type I) macroscopic character of $D$. }

As a conclusion, the full description of $S\cup A \cup D$ with macroscopic quantum states as defined above shows a 
random self-projection of the states of $S$ that are coupled to the states of $D$ {\color{black}through $A$}, 
the latter being chosen to highlight some desired state of $S$. {\color{black}As compared to the usual description of 
measurement through decoherence via loss of information in the environment, it neither {\color{black} requires such} a 
role for the environment, nor does it {\color{black} imply} the creation through a unitary evolution of superposed 
macroscopic states:}  {\color{black}they are forbidden indeed, {\color{black} by} the very structure of the von Neumann algebra. }

\subsection{The room for quantum} 

If the mathematical formalism described in section \ref{sec_infiniteTensorProducts} sheds light on the reason why macroscopic systems built out of many quantum microscopic elements do not behave in a quantum way, this formalism also lets understand where the quantum phenomena have room to exist. In the first parts of this section we have considered large systems with states that differ by a macroscopic fraction of their subsystem states 
i.e. with a number $N_{\neq}$ of subsystems that are in a different state that remains in a finite ratio 
$\xi =N_{\neq}/N $ with $N$ when $N\rightarrow \infty$. Similarly, the Operator sectorisation limit lemma applied to the specific case of unitary evolution operators entails that macroscopic evolution of macroscopic states goes from sector to sector, all different at each time, and  that summing the amplitudes of each microscopic variant of the macroscopic state leaves a probability 1 to be in the sector. Macroscopically, evolution is deterministic. 

{\color{black} However these arguments collapse} if the macroscopic system states belong to the same 
sector, i.e. differ by a finite number $N_{\neq}$ of microscopic states, with $N_{\neq}/N \rightarrow 0$ in the macroscopic limit.
So the state of a macroscopic system can be seen as a superposition only if the superposed state differ by a finite number of microscopic states. As a matter of {\color{black}fact}, opposite to the full state{\color{black}-}space of large systems, the sectors are uncountably many separable sub-spaces.  This is the case of usual quantum experiments if one considers a few (more precisely, a finite number of) quantum elements that interact with a measurement device. In this case, unitary evolution with several accessible states 
remains possible among a finite number of microscopically different states. 

A quantitative statement of how much room there is left for quantum -- i.e. how many subsystems can be considered before entering in the classical realm -- has to be stated carefully.
In principle the radical change of behaviour occurs in the $N\rightarrow\infty$ limit only, 
so do we ``really" need the mathematical infinities in order to consistently include the classical world in the formalism? 

A tentative answer is yes, and this is a major difference between the present approach and others, which claim that {\color{black} measurement contexts} are ``only" large quantum systems.  In opposition, our claim here is that physical 
{\color{black} measurement contexts}  are needed from the beginning to define ``quantum states";  then the impact on the formalism is that  infinities are required in the {\it mathematical} language, that should not be confused with physical reality.

To soften somehow this strong claim, the Vector sectorisation limit lemma  ensures that 
the scalar products that are needed to maintain the quantum properties vanish (exponentially) progressively, as in the usual decoherence approach. This does not remove the need of completing the mathematical framework in the appropriate way, but 
it validates the ``pre-measurement" approaches that are used to calculate e.g. decoherence times and similar quantities.  We illustrate this important point in the example section below.

\section{Examples with  two-state systems} 
\label{sec_examples}

This section presents simple examples  to illustrate the above considerations, and enters in more details about their implications for quantum measurements. 

\subsection{Spins $1/2$ in a magnetic field}

Let us consider a collection of $N$ spin $\frac{1}{2}$ objects, even without interaction between these objects. This can be e.g. the spins of fermionic nuclei in a cold gas, but there is no fundamental difference with the spins of a Heisenberg model. Each spin $i$ is described using a two-dimensional Hilbert space $\mathcal{H}_i$ that is spanned by e.g. the eigenstates of $\hat S^z_i$ that we can note $\vert \uparrow_i\rangle$ and $\vert \downarrow_i\rangle$ or  by the eigenstates of $\hat S^x_i$ that we can note $\vert +_i\rangle$ and $\vert -_i\rangle$. We have 
\begin{equation}
\vert +_i\rangle = \frac{1}{\sqrt{2}}(\vert \uparrow_i\rangle + \vert \downarrow_i\rangle)
\mbox{ and }
\vert -_i\rangle = \frac{1}{\sqrt{2}}(\vert \uparrow_i\rangle - \vert \downarrow_i\rangle)
\end{equation}
The complete space of states is the tensor product of these $N$ two-dimensional 
spaces $\mathcal{H}=\otimes_{i=1}^N\mathcal{H}_i$.

One can use a magnetic field to align all the spins in the positive $z$ direction. The complete state will be 
\vspace{-2 mm} 
\begin{equation}
\vert \Psi_N\rangle = \otimes_{i=1}^N \vert \uparrow_i\rangle
\end{equation}
Alternatively, one can decide to align the spins along the positive $x$ direction. The complete state will be
\vspace{-2 mm} 
\begin{equation}
\vert \Phi_N\rangle = \otimes_{i=1}^N \vert +_i\rangle
\end{equation}
The common sense of quantum physics tells that since each $\hat S_x$ eigenstate is a superposition of $\hat S_z$ eigenstates, then $\vert \Phi_N\rangle$ must have a non-zero projection on $\vert \Psi_N\rangle$. However
\begin{equation}
\langle\Psi_N\vert \Phi_N\rangle = (1/\sqrt{2})^N 
\end{equation}
so when $N\rightarrow\infty$ their scalar product vanishes. 

What happens is that when the Hilbert space breaks down at the limit, these two states end up in two different sectors of $\mathcal{H}$ because they differ by a macroscopic fraction (actually all in this example) of their elementary spin states. Note that if only a fraction $0<\xi<1$ of the spin states were different, the result would be the same. For $\lim_{N\rightarrow\infty}\langle\Psi_N\vert \Phi_N\rangle $ to remain finite, only a finite number of spins may be in different states. In this case, the two systems only differ by a microscopic fraction of their elementary components, in which case they remain in the same sector, and there is room for quantum

This example further shows the way the system goes towards the breakdown before reaching the limit. Actually, $\Vert \langle\Psi_N\vert \Phi_N\rangle\Vert^2 = 2^{-N}$ is the probability to observe $\vert \Psi_N\rangle$ when the system is in state $\vert \Phi_N\rangle$. When $N$ is large enough, this probability is too small to be observable as it would require a non-realistic number of repetitive measurements to show. This illustrates the meaning of the Vector sectorisation limit lemma in quantum physics.

\subsection{QND measurement of a single ion} 
\label{QND}
As formalised by von Neumann \cite{JvN-Grundlagen}, an ideal quantum measurement has two main outcomes (i) it provides a value that is an eigenvalue of the measured observable, and (ii) it prepares the system in the related eigenstate of this observable. In the absence of a free evolution of this observable after the measurement, such an ideal measurement is repeatable, the same measurement carried out on the same system will give the same result, and the system will remain in the eigenstate. Note however that in the case where the measurement is destructive, only (i) occurs and the measurement cannot be repeated. Note also that measurements that involve an observable with degenerate eigenvalues, the result of (ii) can still be partly undetermined \cite{Lueders,PZHCKH}.
 
In practice, excellent approximations of such ideal measurements are provided by so-called Quantum Non Demolition (QND) measurements, that can be performed with both continuous \cite{qnd_pg} and discrete variables \cite{sh_jmr}.  Discussing such a measurement process is interesting in our frame so as to highlight what happens regarding the preparation of the output state and regarding the production of the observable value. As an example,  we give below some details about the experiment of \cite{qnd_wineland} that performs a repeated QND measurement on a two-state system carried by a single ion qubit. Many other similar experiments have been realised more recently, by using gate-based quantum processors. 

The system is a trapped Al$^+$ ion with two states $\vert \uparrow_{\mbox{\tiny Al}^+} \rangle$ and $\vert \downarrow_{\mbox{\tiny Al}^+} \rangle$. Direct spectroscopy of these states is difficult, so one relies on an ancillary system, a trapped Be$^+$ ion with two energy levels $\vert \uparrow_{\mbox{\tiny Be}^+} \rangle$ and $\vert \downarrow_{\mbox{\tiny Be}^+} \rangle$ that are prone to sympathetic laser cooling and state-dependent resonance fluorescence scattering. Both ions have an extra internal state that is used in the manipulation and detection processes. The initial  state is a superposition for the system and a definite state for the ancillary
\begin{equation}
\vert \mbox{in}\rangle = (\alpha \vert \uparrow_{\mbox{\tiny Al}} \rangle + \beta \vert \downarrow_{\mbox{\tiny Al}} \rangle)\otimes \vert \downarrow_{\mbox{\tiny Be}} \rangle 
\end{equation}
A two step $\pi$-pulse manipulation allows transfering the $\vert \downarrow_{\mbox{\tiny Al}^+} \rangle$ spin state of the Al$^+$ ion to its motion modes and then transfer it back to the Be$^+$ ion spin, leading to complete state
\begin{equation}
\vert \mbox{ent}\rangle = \alpha \vert \downarrow_{\mbox{\tiny Al}} \rangle\otimes \vert \uparrow_{\mbox{\tiny Be}} \rangle + \beta \vert \uparrow_{\mbox{\tiny Al}} \rangle\otimes \vert \downarrow_{\mbox{\tiny Be}} \rangle 
\end{equation}
Then the state of the Be$^+$ ion is probed with a laser to activate a state-dependent resonance fluorescence phenomenon, by which the Beryllium emits $F$ photons $\gamma_i$ if it is in the $\vert \downarrow_{\mbox{\tiny Be}} \rangle$ state and no photon if not. These photons are collected by an intensified CCD where the following sequence of phenomena occurs.

\noindent $\bullet$ some fluorescence photons arrive on a photocathode where each photon triggers the emission of $P \le F$ electrons by photoelectric effect. At this point of the discussion, let us assume the ideal case $P=F$ as $F$ is the relevant value for the {\color{black} size} of the Hilbert space. This amounts to no photon loss, 100\% quantum efficiency and no dark count.

\noindent $\bullet$ each of photoelectrons $e^-_i, i=1,...,F$ is accelerated by an electric field and hits a microchannel plate where it releases each a large number $S_i$ ($i=1,...,F$ with our ideal assumptions) of secondary electrons in a variety of momentum states $\vert {\bf p}_{i,j}\rangle, j=1,...,S_i$ in the ionisation continuum thanks to the acquired energy. 

\noindent $\bullet$ each secondary electron $e^-_{i,j}, j=1,...,S_i$ is again accelerated by an electric field and hits a phosphorescent plate where it generates an even larger number $K_{i,j}$ of photons. 

\noindent $\bullet$ finally each of the $\sum_{i,j} K_{i,j}$ phosphorescence photons (with the same ideal assumptions) enters the CCD camera proper, where it triggers the creation of electron-hole pairs, again in a variety of conduction-band states $\vert {\bf q}_{i,j,k}\rangle$, leading to a final (quite idealised !) state
\begin{multline*}
\vert \mbox{out}\rangle = 
\alpha \vert \downarrow_{\mbox{\tiny Al}^+} \rangle
\otimes_{I=1}^F \otimes_{j=1}^{S_i} \otimes_{k=1}^{K_{i,j}}\vert 0 \rangle \\
+ \beta \vert \uparrow_{\mbox{\tiny Al}^+} \rangle
\otimes_{I=1}^F \otimes_{j=1}^{S_i} \otimes_{k=1}^{K_{i,j}}\vert {\bf q}_{i,j,k}\rangle 
\end{multline*}
and collecting the electrons of the pairs will allow converting their charge in a voltage  signal by capacitor effect. Three aspects are relevant in this process. 

\noindent $\bullet$ First, there is a deterministic and reversible chain of phenomena that leads the (Al$^+$ system $\cup$  device) from state $\vert \mbox{in}\rangle$ to state $\vert \mbox{ent}\rangle$. The Hilbert space size increases slightly as a few new degrees of freedom are added. This basically propagates entanglement to the ancilla, in a so-called ``pre-measurement". 

\noindent $\bullet$ Second, the phenomena that take state $\vert \mbox{ent}\rangle$ to state $\vert \mbox{out}\rangle$ rely on a drastic increase of the number of involved degrees of freedom, that is finally macroscopic so as to be measured as a voltage. In this stage, the size of the Hilbert space increases by involving so many tensor products that the two states 
\begin{equation} 
\otimes_{I=1}^F \otimes_{j=1}^{S_i} \otimes_{k=1}^{K_{i,j}}\vert 0 \rangle
\mbox{ and } 
\otimes_{I=1}^F \otimes_{j=1}^{S_i} \otimes_{k=1}^{K_{i,j}}\vert {\bf q}_{i,j,k}\rangle
\end{equation}
can be considered as belonging to two different sectors of the resulting large Hilbert space, so orthogonal (up to the Vector sectorisation limit lemma) and thus the density matrix $ \vert \mbox{out}\rangle \langle\mbox{out}\vert $ is sector diagonal.

Even in the most idealised case, we are thus in a classical statistical mixture, where quantum correlations have been irreversibly lost, because once initiated this process brings the result to the (unbounded) macroscopic level. It is worth noting that 
in this example an avalanche phenomenon allows an exponential growth of the involved quantum degree of freedom number, thus getting macroscopic in a finite time and energy.
The completion of the measurement process will 
select randomly one of the possible issues in the density matrix and complete the projection to either state. If one does not take the ideal assumptions, introducing losses, quantum efficiency below 100\% and dark count electrons, one has to trace out the lost degrees of freedom and add noise, in which case the irreversible loss of information is even more obvious, even though these effects are not strictly necessary. 

\noindent $\bullet$ Third, one sees that while the states of the various ancillas are destroyed in the measurement process, the state of the Al$^+$ system is 
projected with probability $\vert \alpha\vert ^2$ to state $\vert \downarrow_{\mbox{\tiny Al}} \rangle$ and probability $\vert \beta\vert ^2$ to state $\vert \uparrow_{\mbox{\tiny Al}} \rangle$. In particular, if the process is carried out a second time on the system, it will lead to the same result (up to measurement errors). This is the QND aspect of this example.

To conclude this discussion, it is clear that the framework of standard QM {\color{black} works not by answering the points raised in the introduction, but simply by taking them for granted:} there is clearly a classical experimental setup, involving traps, lasers, photodetectors, much larger  than the quantum system under consideration, and at this level the measurement results are
fundamentally random \cite{random}. The cut is not at the level of the entanglement between the system and the ancilla, that is reversible, but at the level of the identification of the ancilla's state, that brings the result at the ``unbounded" macroscopic level.  Since this level is not included in the standard (Copenhagen) formalism, we are unable to tell where the system ends, and where the measurement context begins: this is the usual ``quantum measurement problem". 

So in order to fully exploit the extended formalism of the previous sections, we move now to a different point of view, provided by the framework of Contexts, Systems and Modalities (CSM). 

\section{Discussion in the CSM framework.}
\label{sec_CSM_wrap_up}

\subsection{The CSM approach}

Contexts-Systems-Modalities (CSM) is an approach developed in the recent years to provide an objective presentation of quantum mechanics \cite{CO2002,csm1,random,generic,csm4b,csm4c,grangierCompleting,ContextualInference,myst,context}. It allows one to derive the usual QM  formalism, in a form related to the usual textbook approach,  
while clarifying the role of the various objects used in the formalism, and 
simplifying thus a number of interpretation issues. Our purpose here is to show that the elements presented in the previous sections can complete and close the CSM conceptual framework; we will do so by developing ideas introduced  in \cite{grangierCompleting}.

According to CSM, objectivity is  associated with the possibility to define measurements results that are repeatable, and can be predicted with certainty \cite{CO2002}. As discussed in the previous section, this is quite possible in QM, simply by repeating the same measurement on the same system, using the same (idealised)  apparatus, called a context. Such a repeatable result is called a modality, and it belongs to a system within a context.

Again according to CSM, the fundamental difference between classical and quantum physics is that modalities are both quantised and contextual, in a sense defined in 
\cite{myst,context}. 
It is also understood that contexts are self-contextual, i.e. a  built-in Heisenberg cut is initially postulated. However  a legitimate question is to ask how such a postulate can be made consistent with the whole QM mathematical formalism -- this is the issue we want to address now.

In the framework of the present paper, the properties of macroscopic quantum systems  (including contexts) derived in Section \ref{sec_largeSystems} can explain how the Heisenberg cut can be embedded in the formalism. 
In short, even though built out of many quantum elements, contexts are large enough to be in one sector of the tensor product of their constituent Hilbert spaces, and therefore in a state where no superposition is possible; then the outcomes of measurements on a context do not depend on the conditions in which they are carried out. In other words, as we will discuss in details below, macroscopic systems are non-contextual, and thus can be described as classical physical objects. 

Following  \cite{grangierCompleting}, we will also use some notions from operator algebras, that originate from articles written by Murray and von Neumann in the 1930's \cite{OpA}.  In standard textbook,  QM uses only the so-called type-I (von Neumann) algebras, associated with a finite number of particles,  
where Hilbert spaces are separable, with a finite or countably infinite dimension. Since an infinite number of particles is considered here, we have to include also  type-II or type-III algebras, that are physically less usual, and mathematically more sophisticated but still well understood  \cite{grangierCompleting,OpA,AQFT,vitiello}.

\subsection{Defining systems, contexts and modalities} 

A basic CSM idea is to split the whole physical world into a (bounded) subpart that will be considered as the system, and a much larger part, modeled as unbounded, considered as the context and obeying the rules  described in the previous sections.  By definition a modality is obtained from a measurement, and then it is repeatable with certainty, as long as the system and the context are kept the same (repeatability is strict in the absence of free evolution, otherwise it is replaced by full predictability in another context). Therefore a modality can be seen either as the result of a measurement, or as the ``identity card" \cite{random} associated with the repeatability  of this result. 
This is why it corresponds to a ``state of a given system within a given context", even when the system and the context are not interacting any more. Spelling out the context where the modality has been defined is a specific feature of CSM, differing from the Copenhagen point of view; as a consequence, the usual $\vert \psi \rangle$ is not associated with a modality, but with an equivalence class of modalities belonging to different contexts, called an extravalence class \cite{csm4b}. Another difference is that CSM has a well-defined -- non-classical -- ontology, see Annex A2.

When changing the context, repeatability does not hold any more, and  observed modalities in the new context are in general random. This randomness appears to be key to save causality in EPR or teleportation experiments, see e.g. \cite{ContextualInference}, imposing a non-classical probabilistic treatment \cite{svozil}. 
Then an important advantage of CSM is  that, still without introducing the dynamics of the measurement, it allows one to justify physically Gleason's hypotheses to compute these probabilities. Therefore Born's rule can be obtained from Gleason's theorem \cite{GleasonTheorem,csm4b,csm4c}, showing clearly that the probability rule comes only from the geometry of the (projective) description of probabilities, without requiring any dynamical elements. In this point of view the modalities are context-dependent and asymptotic, i.e. they are defined ``long before" or ``long after'' the measurements \cite{grangierCompleting}. By doing so, the usual QM formalism is recovered, 
but with significant conceptual differences \cite{myst}. 

\subsection{``Actualizing" the measurement} 

Then the question arises: how to describe the measurement itself, i.e. the change from some modality  in a context to another one in another context ? Clearly the dynamics should involve additional algebraic elements or projectors, but they cannot be built by ``superposition" of the available ones, due to the sectorisation. 

In the present framework,  the dynamics that connects mathematically the asymptotic states 
must involve time evolution within a type III algebra, with a fundamentally random element to avoid introducing hidden variables. The machinery of this dynamics remains to be 
explicited, but many parts are already known: the ``kinematics" defined by the initial and final modalities, the probability rule connecting them given by Born's law, and the way to approximate ``what is going on", using decoherence-type calculation {\color{black} during the pre-measurement step (see below).} Only the final jump to a new ``actualised" modality is missing, because it requires non-standard mathematical tools, that are made necessary from the previous discussion \cite{grangierCompleting}. 

On the physical side, how to deal with this problem is well known in practice:  the measurement can be described in an approximate way, by considering  the apparatus as a large collection of ``ancillas" coupled with the initial system, and getting entangled with it, 
{\color{black} as detailed in the previous sections. It is essential to remember that such a large ``pre-measuring device"} is always embedded in a context, where the system's asymptotic modalities are already identified, from  the very definition of the quantum system's Hilbert space and the choice of the measurement device. Therefore a well-defined modality must be recovered at the end of the process, if it has any meaning as a quantum measurement on the isolated system of interest. 

But then, as long as the measurement does not ``really happen'' as a proper quantum jump (i.e., reaching a new asymptotic modality), the coupling with more and more ancillas may be described as a unitary evolution of the initial modality, respecting the algebraic sectorisation.  
These evolutions are in principle reversible, including entanglement and dis-entanglement with ancillas, and the effect of external Hamiltonian like a classical magnetic field acting on a spin. When the measurement really happens, associated with irreversibility \cite{grangierCompleting,drossel,ThermodynLimit}, the global dynamical evolution makes it impossible to maintain a separation between any well-defined system and the context, and therefore the very conditions for a unitary evolution of the system are lost. But it is known in advance that the system and context will end up in another modality, as prescribed by Born's rule, and imposed by the whole kinematics structure resulting from the initial definition of the system.
In summary: 

- the post-measurement behaviour of the system depends only on the (type~I) system's projectors and Born's rule, whereas the (type~III) context's projectors are needed to make the mathematical description of the modality objective and complete;

- the extended commutative algebra including all projectors is not type~I, because the context's projectors are operating in a space with non-countable infinite dimension. Crucially, this space is ``sectorised", as it is usual in  non-type~I algebras \cite{AQFT,vitiello}. Physically this corresponds to the fact that  a unique ``context-and-result" is defined outside measurement periods;

-  on the other hand, the non-commuting observables of the system act in a separable Hilbert  space with either finite or countably infinite dimension (type~I). Formally this is just usual QM, but dressed in a quite different way. 

\subsection{Looking again at the ``Heisenberg cut"} 

From the above it also becomes clear that the ``cut" between the system and the context  is simply a consequence of the very definition of an isolated system.  As mentioned before, a quantum isolated system differs fundamentally from a classical one, because its physical properties (modalities) are quantised in a given context; therefore it is impossible to ignore this context, as it would be the case in classical physics. 
The cut is moveable depending on the definition of the system, but in order to properly define a closed quantum system, it must  fulfill well-defined conditions: 
\begin{itemize}
\item there is only one cut, because by construction  the context's projectors act in an unbounded space. It is however possible to increase the size of the system and to redefine its dimension 
as long as it remains countable,  moving the cut accordingly;
\item in order to get a consistent unitary evolution of the system's  projectors, the choice for the cut,  i.e. the  definition of the boundary between system and context, must be agreed upon a priori,  following the definition of the system;
\item there is no restriction on the size of the system, but it must always be embedded in a context, in agreement with the two previous rules. Correspondingly, there is always a context ``at the infinity edge". 
\end{itemize}

Let us emphasise again that, just like usual QM, we obtain 
the quantum description of isolated systems within contexts, {\it outside} of measurements periods, though measurements may be approached as close as we wish for all practical purposes (``FAPP"). The full non-separable Hilbert space is required for mathematical consistency during the unbounded part of the measurement (see also Annex A2), but outside of this,  one can simply use the standard quantum formalism, building a dynamical model of the system-context interactions, and making suitable approximations. Then it can be shown \cite{ThermodynLimit} that ``the pointer observables behave as if they commuted with any observable of the algebra; however, this holds only in the final states (i.e., when the measurement is fully completed)". This shows the consistency between  the present approach, starting from general physical principles and implementing them within the algebraic formalism, and a more dynamical approach, starting from usual QM and making suitable approximations. In both cases,  there is only one ``factual state" of the macroscopic world, and the ``full quantum state" (or modality) of a subsystem  belongs to a quantum system within  a classical context. 
\section{Conclusion}
In order to bring a new view on the possible quantum behaviour of large quantum systems, we have explored the consequences of a mathematical article by John von Neumann \cite{vonNeumann1939} dedicated to infinite tensor products. The main  mathematical result of \cite{vonNeumann1939}, physically useful for our goal, is that  the non-separable infinite tensor product of many Hilbert spaces breaks down into a direct sum of  infinitely many, separable orthogonal sectors.

The properties of macroscopic quantum systems are derived from the theorems of von Neumann, by considering as in statistical physics that the limit where the number of elementary subsystems $N\rightarrow \infty$ describes the macroscopic behaviour of a macroscopic system.  In the quantum framework, the breakdown theorem means that the states of macroscopic systems that differ macroscopically from one another will be orthogonal, allowing no superposition of macroscopically distinct states, in the sense spelled out above. This leads to a built-in von Neumann decoherence, that is not ``environment-induced" since the unbounded context / environment is already an intrinsic part of the mathematical description. 

{\color{black} This sheds a new light on t}he measurement process can be clarified in this vision, by considering system and detector as a macroscopic quantum object. 
{\color{black} This points to the need of a revised description of  this process, that has to involve a} fundamental randomness, imposed by contextual quantisation; this looks  similar to statistical physics, but  makes sense {\it only}  when the context has been actualised. We highlight that this randomness has a profound link with causality: actually, it allows the quantum formalism to preserve relativistic causality even in the presence of a ``global" contextuality, like it appears in Bell-type scenarios \cite{ContextualInference}.

It should be clear that in this article we are dealing with idealised textbook QM, where the very idea of an isolated quantum system makes sense. In the more general framework of open quantum systems, the system (e.g. an atom) is not fully isolated, but weakly coupled to a reservoir (e.g. the quantised electromagnetic field), and the evolution of the system alone is not unitary. Such a situation can be included in the present framework, as it is included in standard quantum field theory; what really matters for consistency  is that there is always a context ``at the infinity edge".

Finally, by restating these considerations in the framework of Contexts, Systems and Modalities, their application to the limit of large systems allows us to verify the complete consistency of the approach.
Classical contexts are postulated to begin with, then through Gleason and Uhlhorn the usual Hilbert space QM formalism is derived, and when pushed to the macroscopic limit, it provides the quantum description of the classical contexts, consistently closing the loop. This is why the present approach can be viewed as a {\it contextual} unification of classical and quantum physics.
\vskip 12mm

{\bf Acknowledgments: } 
PG thanks Alexia Auff\`eves, Nayla Farouki, Franck Lalo\"e, Roger Balian, Karl Svozil, for many interesting and useful discussions. MvdB thanks the Thales engineers discovering quantum technologies for their candid but actually extremely relevant questions.

\section{Annex}

\section*{A1 Proof of the sectorisation limit lemmata}\label{secA1}

\label{sec_proofs}

\subsection*{Proof of the Vector sectorisation limit lemma}
\noindent Let $\mathcal{C}$ and $\mathcal{C}'$ be two distinct sectors of $\mathcal{H} = \otimes_{\alpha\in I}\mathcal{H}_\alpha$.

\noindent Let $\mathcal{C} \ni \vert \Phi\rangle = \otimes_{\alpha \in I} \vert \phi_\alpha\rangle$ and $\mathcal{C}'\ni\vert \Psi\rangle = \otimes_{\alpha \in I} \vert \psi_\alpha\rangle$ .

\noindent  By Breakdown theorem a),  one has
$\langle \Psi\vert \Phi\rangle = \prod_{\alpha \in I} \langle \psi_\alpha\vert  \phi_\alpha\rangle=0$
\vskip 2mm

\noindent  By the definition of the convergence of infinite products, the fact that this infinite complex product converges to value 0 means that 
$ \forall \varepsilon>0, 
\exists \mbox{ a finite } J_\varepsilon\subset I$  such that $\forall I_N =\{ \alpha_1, ...,\alpha_N\}\subset I$, all $\alpha_i$'s different and $  
J_\varepsilon \subset I_N$, and then 
$\vert \prod_{\alpha \in I_N} \langle \psi_\alpha\vert  \phi_\alpha\rangle\vert  < \varepsilon.$
Thereby  $\vert \Phi_N\rangle = \otimes_{\alpha \in I_N} \vert \phi_\alpha\rangle$ and $\vert \Psi_N\rangle = \otimes_{\alpha \in I_N} \vert \psi_\alpha\rangle$ $\Box$.

\subsection*{Proof of the Operator sectorisation limit lemma}

{\color{black}As $\vert \Psi\rangle \in \mathcal{C} \ni \vert \Phi\rangle$, $\#\{\alpha\in I, \vert \psi_\alpha \rangle \neq \vert \phi_\alpha\rangle\}\l \infty$.}
As {\color{black}$\hat A$} is defined over $\mathcal{H}$, and $\mathcal{H}$ being the direct sum of all sectors, including $\mathcal{C}$, by the principle of the excluded middle, $\forall \vert \Psi\rangle \in \mathcal{C}$, either ${\color{black}\hat A}  \vert \Psi\rangle \in \mathcal{C}$ or ${\color{black}\hat A}  \vert \Psi\rangle \notin \mathcal{C}$.

\noindent $\bullet$ $ {\color{black}\hat A}  \vert \Psi\rangle \in \mathcal{C} \Leftrightarrow {\color{black}\hat A}$ has non-zero matrix elements only for 
$ (\vert \psi_\alpha\rangle,\vert \phi_\alpha\rangle ), \alpha\in I_N:=\{\alpha_1,...,\alpha_N \}$ finite {\color{black}, i.e. $p(\hat A)\l \infty$}. 
In this case, ${\color{black}\hat A}$ is of the form 
${\color{black}\hat A} = {\color{black}\hat A}_N \otimes \otimes_{\alpha \in I\setminus I_N }\hat I_\alpha$, 
where ${\color{black}\hat A}_N$ can be viewed as an $N\times N$ complex matrix that allows  computing the inner product $ \langle {\color{black}\Phi}\vert  {\color{black}\hat A}  \vert \Psi\rangle$ to a finite complex value. 

\noindent $\bullet$ If ${\color{black}\hat A}$ acts on an infinite number of $\alpha_i$'s {\color{black}i.e. $p(\hat A)$ infinite}, one still has to prove that ${\color{black}\hat A}  \vert \Psi\rangle$ relies on convergent sequences. For this let us examine $\hat A  \vert \Psi\rangle$. Since  $\vert \Psi \rangle \in \mathcal{H}$
\begin{equation}
\vert \Psi\rangle = \sum_{k \in K} {\mathit \Psi_k} \otimes_{\alpha \in I} \vert \psi_\alpha^k\rangle
\end{equation}
where the $\{ \vert \psi_\alpha^k\rangle\}_{\alpha \in I}, k\in K$ are $\# K$ convergent sequences. Now ${\color{black}\hat A}$ is not necessarily the tensor product of operators in each $\mathcal{H}_{\color{black}\alpha}$, but at least shall decompose on the sum of such products, 
\begin{equation}
{\color{black}\hat A} = \sum_{p} {\color{black}a}_p \otimes_{\alpha \in I} {\color{black}\hat A}^p_\alpha
\mbox{ and }
{\color{black}\hat A}\vert \Psi\rangle = \sum_{k,p} {\mathit \Psi_k} {\color{black}a}_p\otimes_{\alpha \in I} {\color{black}\hat A}^p_\alpha\vert \psi_\alpha^k\rangle.
\end{equation}
\vspace{-2mm}
So
\vspace{-2mm}
\begin{equation}
\langle \Psi \vert  {\color{black}\hat A}^\dagger {\color{black}\hat A}\vert  \Psi \rangle =
\sum_{k,l} \mathit{\Psi_k^*\Psi_l}\sum_{p,q}{\color{black}a}_p^*{\color{black}a}_q{\color{black}\prod_{\alpha \in I}}\langle \psi_\alpha^k\vert {\color{black}\hat A}_\alpha^{p \dagger}{\color{black}\hat A}_\alpha^q \vert \psi_\alpha^l\rangle
\end{equation} 
and
$  \langle \Psi \vert  \Psi \rangle = \sum_{k,l}\mathit{\Psi_k^*\Psi_l}$

Now, as ${\color{black}\hat A}$ is bounded, $\exists c\in\mathbb{R}, \langle \Psi \vert   {\color{black}\hat A}^\dagger {\color{black}\hat A} \vert  \Psi \rangle \le c \langle \Psi \vert  \Psi \rangle$. As $  \langle \Psi \vert  \Psi \rangle $ is bounded so is $\langle \Psi \vert  {\color{black}\hat A}^\dagger {\color{black}\hat A} \vert  \Psi \rangle $ which entails that ${\color{black}\hat A}  \vert \Psi\rangle$ relies on convergent sequences $\{{\color{black}\hat A}_\alpha^q \vert \psi_\alpha^l\rangle\}_{\alpha \in I}, \forall q,l$. Therefore ${\color{black}\hat A}  \vert \Psi\rangle \in \mathcal{H \setminus C}$ so the Vector sectorisation limit lemma applies {\color{black}to $\langle \Phi \vert \hat A \vert \Psi\rangle.$}$\Box$. 

\vspace{-3mm}
\section*{A2 About ontology and infinities. }
The CSM ontology relies on the general idea of contextual objectivity, as introduced in \cite{CO2002}. In this framework the systems within contexts are objectively real, modalities are actual  (repeatable) events, and projectors in Hilbert spaces are mathematical tools used to calculate non-classical probabilities. For CSM there is only one factual reality, the role of agents is the same as in classical probabilities, and there are no hidden variables -- but the usual state vector $\vert  \psi \rangle$ must be completed by specifying the measurement/observation context in order to define a modality. The so-called quantum non-locality is better understood as predictive incompleteness, associated with the fundamental randomness that is a necessary consequence of contextual quantisation  and is fully compatible with relativistic causality \cite{ContextualInference}.  All these features fit well in the mathematical framework presented in this article, and since there is no free lunch, the price to pay is embedding infinities in the formalism.

A very interesting discussion on the role of infinities in physics and mathematics can be found in a conference given in 1925 by David Hilbert \cite{Hilbert}. On the one hand Hilbert writes that infinities (as defined by Cantor) are a required ingredient for the completeness on mathematics: ``No one shall drive us out of the paradise which Cantor has created for us". On the other hand he writes that ``The infinite is nowhere to be found in reality, no matter what experiences, observations, and knowledge are appealed to". Our point of view is somehow different: we consider that, with respect to natural sciences,  mathematics are basically a language used to describe (or to model) physical reality: they are neither  reality itself, nor are we constrained to any sort of ontological identity between physics and mathematics. 

Admitting that, there is no reason for not using the full power of the mathematical language, including infinities. 
In practice this is done already in differential or integral calculus; but here these ideas are pushed further, towards Cantor-type incommensurability, with  not only numbers but full algebraic structures appearing at the infinite limit. 
It is true that such concepts are not easy to grasp, and may lead to currently intractable calculations; but this should not be a reason for the physicists not to enter Cantor's paradise. 


\end{document}